\pdfoutput=1
\documentclass[a4paper]{article}

\usepackage{INTERSPEECH2020}
\usepackage{mathtools}
\usepackage{multirow}
\usepackage{mathrsfs}
\usepackage[a-2b,mathxmp]{pdfx}[2018/12/22]

\title{They are wearing a mask! Identification of Subjects Wearing a Surgical Mask from their Speech by means of x-vectors and Fisher Vectors}
\name{Jos\'e Vicente Egas-L\'opez$^1$}
\address{
$^1$Institute of Informatics, University of Szeged, Szeged, Hungary}
\email{egasj @ inf.u-szeged.edu}

\begin{document}

\maketitle
\begin{abstract} 
Challenges based on Computational Paralinguistics in the INTERSPEECH Conference have always had a good reception among the attendees, owing to its competitive academic and research demands. This year, the INTERSPEECH 2020 Computational Paralinguistics Challenge offers three different problems; here, the {\it Mask} Sub-Challenge is of specific interest. This challenge involves the classification of speech recorded from subjects while wearing a surgical mask. In this study, to address the above mentioned problem we employ two different types of feature extraction methods. The x-vectors embeddings, which is the current state-of-the-art approach for Speaker Recognition; and the Fisher Vector (FV), that is a method originally intended for Image Recognition, but here we utilize it to discriminate utterances. These approaches employ distinct frame-level representations: MFCC and PLP. Using Support Vector Machines (SVM) as the classifier, we perform a technical comparison between the performances of the FV encodings and the x-vector embeddings for this particular classification task.
We find that the Fisher vector encodings provide better representations of the utterances than the x-vectors do for this specific dataset. Moreover, we show that a fusion of our best configurations outperforms all the baseline scores of the \textit{Mask} Sub-Challenge.
\end{abstract}

\section{Introduction}
The Computational Paralinguistics differs from Automatic Speech Recognition in that the latter seeks to determine the {\it content} of the speech of an utterance, while the former seeks to understand the {\it way} that the speech is spoken. There are different types of techniques that attempt to solve this problem in Computational Paralinguistics. Methods such as the i-vector Approach, the Fisher vector, neural networks, among others, 
are being increasingly used by researchers to address paralinguistic issues.  This can be seen in studies like diagnosing neurodegenerative diseases using the speech of the patients~\cite{egas2019park,gosztolya2018identifying,grosz2015assessing}; the discrimination of crying sounds and heartbeats~\cite{gosztolya2018generalutterancelevel}; or the estimation of the sincerity of apologies~\cite{gosztolya2016estimating}. These studies aim to distinguish the latent patterns existing within the speech of a subject and not the content of it.

The INTERSPEECH ComParE Challenge, annually organized since 2009~\cite{schuller2011recognising}, has provided a wide variety of Computational Paralinguistics problems each year. These types of challenges seem to encourage its participants to use or devise state-of-the-art techniques to handle the states and characteristics latent in an audio signal. This year, the challenge offers three tasks; but here we will just focus on one of them, namely, the \textit{Mask} Sub-Challenge. 

The above-mentioned challenge involves the following: speakers (i.e. German natives) were recorded while wearing a surgical mask, and also while not wearing one. The task is to determine whether the utterance corresponds to a speaker whose speech was recorded while wearing the mask or not. The baseline reported by the organizers is a UAR (Unweighted Average Recall) score of 70.8\%, which corresponds to a non-fused score. And a 71.8\% for the fusion of the best four configurations for the \textit{Mask} Sub-Challenge.
Forensics and \textit{'live'} communication between surgeons may benefit from a system that could determine whether a subject is wearing a mask based on their speech~\cite{compare2020}. 

Lots of speaker recognition systems these days are based on i-vectors~\cite{frontendivec}. The i-vector system utilizes a GMM-UBM (Universal Background Model) to extract a fixed-dimension feature called \textit{i-vector}. This is a robust technique that was and still is the state-of-the-art for many speaker recognition/verification approaches~\cite{ibrahim2018vector,garcia2011analysis}. Also, i-vectors have been used in computational paralinguistics and offer promising results when assessing Alzheimer's from speech~\cite{lopez2019alz}, or at the moment of classifying depressed speech~\cite{nicholasivecs2014depr}. Nonetheless, there are more meaningful features that seem to provide better representations of frame-level features than the the i-vectors do. 

Embeddings extracted from a Feed-Forward Deep Neural Network are gradually replacing i-vectors; such embeddings are called \textit{x-vectors}. Regarded as the new state-of-the-art technique for speaker recognition systems~\cite{snyderxvecs}, x-vectors can capture meta-information such as the gender of the speaker, as well as their speech rate (i.e. long-term speech traits). 
Researchers are increasingly using such representations in their studies, especially in text-independent approaches (see e.g.~\cite{silnova2018fast,snydermultispk,chung2018voxceleb2,novotny2018use}). Also, \textit{x-vectors} have already been applied to paralinguistics; studies like~\cite{zargarbashi2019multi,pappagari2020x,raj2019probing} reported high performances at classifying emotions, Alzheimer's, or age and gender of subjects.

As a contribution to the ComParE Challenge, here, we perform the chosen task via two different methodologies. The Fisher Vector (FV) approach~\cite{jaakkola1998exploiting}, which is an encoding method originally developed to represent images as gradients of a global generative GMM of low-level image descriptors; mainly used in image recognition~\cite{song2017adapting}. And we also employ the DNN embeddings approach (i.e. \textit{x-vector system}) where the role of the DNN is to perform a mapping between variable-length utterances and fixed-dimensional embeddings.

The workflow proposed is the following. First, we use two types of frame-level representations, i.e., MFCCs and PLPs extracted from the audio signals. Second, we process the frame-level information obtained utilizing two different techniques: the FV and the x-vector approaches. And third, we classify and evaluate FV and x-vector features individually. Finally, we opt for a late-fusion of the best configurations.

\section{Data}
The Mask Augsburg Speech Corpus (MASC) comprises recordings of 32 German native speakers. It has a total duration of 10 h 9 min 14 sec; segmented into chunks of 1 sec. The recordings have a rate of 16 kHz. The total number of utterances is 36554. The subjects were asked to perform specific types of tasks and recorded their speech while wearing and not wearing a surgical mask. Following the guidelines of the organizers of the Challenge, here, we describe the dataset just briefly (see~\cite{compare2020}).

\begin{figure*}[t!]
\includegraphics[width=0.8\textwidth]{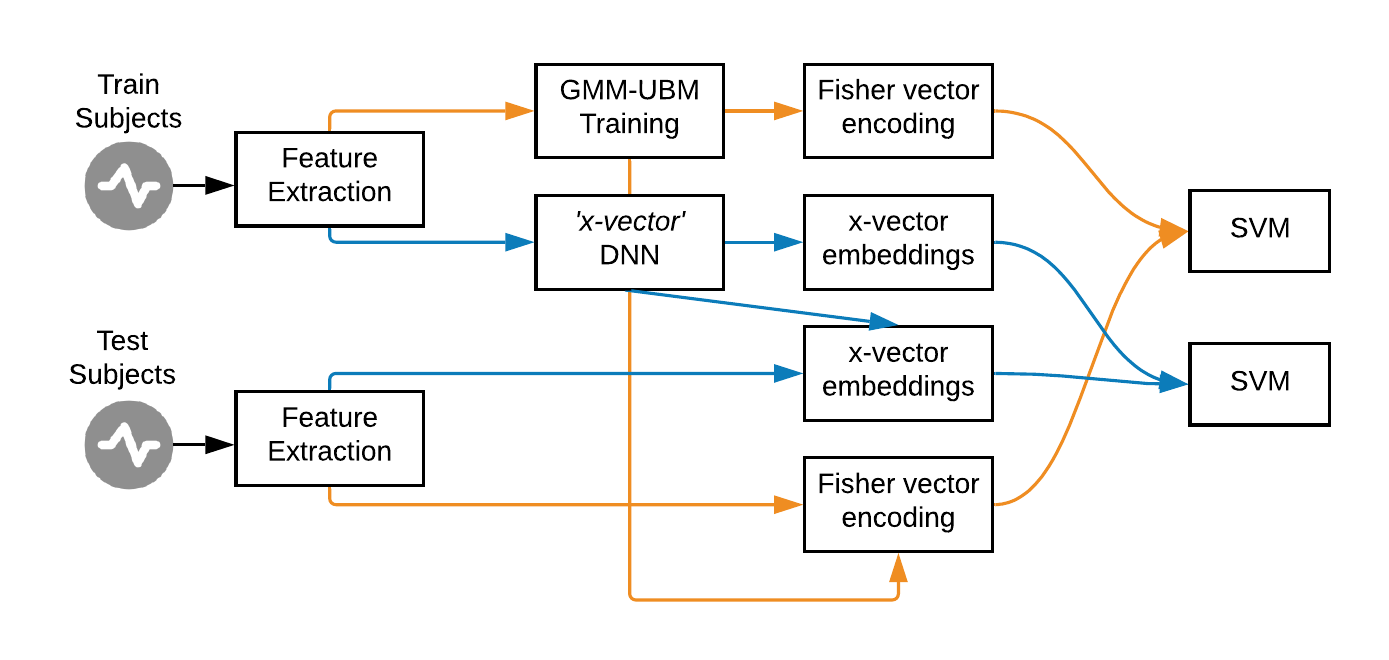}
\centering
\caption{The generic methodology applied in this study.}
\label{fig_syst}
\end{figure*}

\section{Feature Extraction and Evaluation Methods}
As depicted in Figure~\ref{fig_syst}, the steps carried out in our study are as follows: (1) Feature extraction (MFCCs and PLP); (2a) Train GMM-UBM using utterances from the training set, (2b) Train the DNN for the x-vectors utilizing the training set and its \textit{augmented} version; (3a) Extract Fisher vector features from the datasets employing the GMM-UBM model, (3b) Extract embeddings from the DNN; and, (4a/4b) Independently classify the FV and x-vectors representations using SVM. 

\subsection{Frame-level features}
Here, we used the well-known Mel-Frequency Cepstral Coefficients (MFCC) and Perceptual Linear Predictions (PLP) frame-level representations. Both have 13 dimensions, a frame-length of 25ms and a sliding window of 3ms. 
Moreover, since x-vectors are extracted from a DNN, an additional configuration for the MFCCs called \textit{high-resolution (hires)} was utilized. This allows us to maintain all the cepstra while decorrelating the MFCCs. The MFCC-hires configuration is intended for neural network training. This configuration has the same values as those previously described, except that it extracts 40 cepstral coefficients, the number of mel-bins is 40, and the low and high cut-off frequencies are 20 and -400, respectively (see e.g. in~\cite{hernandez2018ted}).
Also, non-speech frames are removed from all the representations employing VAD.

\begin{table}[!t]
\caption{DNN architecture of the x-vector system. It comprises five frame-level layers, a statistics pooling layer, two segment-layers and a final softmax layer as output. $N$ represents the number of training speakers in the softmax layer. The DNN structure here is the same as that given in Snyder et al.~\cite{Snyder2017DeepNN}.}
\begin{center} \label{table_dnn}
\begin{tabular}{@{}cccc@{}}
\toprule
\textbf{Layer} & \textbf{Layer context}  & \textbf{Tot. context} & \textbf{In, Out} \\ \midrule
frame1         & {[}t-2, t+2{]}                 &5                  & 120, 512               \\
frame2         & \{t-2, t,  t+2\}               &9                  & 1536, 512              \\
frame3         & \{t-3, t, t+3\}                &15                 & 1536, 512              \\
frame4         & \{t\}                          &15                 & 512, 512               \\
frame5         & \{t\}                          &15                 & 512, 1500              \\
stats pooling  & {[}0, T\}                      &T                  & 1500T, 3000            \\
segment6       & \{0\}                          &T                  & 3000, 512              \\
segment7       & \{0\}                          &T                  & 512, 512               \\
softmax        & \{0\}                          &T                  & 512, N                 \\ \bottomrule
\end{tabular}
\end{center}
\end{table}

\subsection{x-vectors}
The x-vector approach can be thought as of a neural network feature extraction technique that provides fixed-dimensional embeddings corresponding to variable-length utterances. Such a system can be viewed as a feed-forward Deep Neural Network (DNN) that computes such embeddings. Below, we will describe the architecture of the DNN (based on~\cite{snyderxvecs}) and the embeddings that are extracted from it.

\subsubsection{DNN structure}
Table~\ref{table_dnn} outlines the architecture of the DNN. The \textit{frame-level} layers have a time-delay architecture, and let us assume that $t$ is the actual time step. At the input, the frames are spliced together; namely, the input to the current layer is the spliced output of the previous layer (i.e. input to layer \textit{frame3} is the spliced output of layer \textit{frame2}, at frames $t-3$ and $t+3$). Next, the \textit{stats pooling} layer gets the $T$ frame-level output from the last frame-level layer (\textit{frame5}), aggregates over the input segment, and computes the mean and standard deviation. The mean and the standard deviation are concatenated and used as input for the next \textit{segment} layers; from any of these layers the \textit{x-vectors} embeddings can be extracted. And finally, the \textit{softmax} output layer (which is discarded after training the DNN)~\cite{snyderxvecs, Snyder2017DeepNN, snyderendtoend2016}.

Instead of predicting frames, the network is trained to predict speakers from variable-length utterances. Namely, it is trained to classify the $N$ speakers present in the train set utilizing a multi-class cross entropy objective function (see Eq.~\ref{eq_entropy}). Let $K$ be the number of speakers in $N$ training segmen\-ts. Then, the probability of the speaker $k$ given $T$ input frames $(x_{1}^{(n)}, x_{2}^{(n)}, ...,x_{1:T}^{(n)})$ is given by: $P(spkr_k | x_{1:T}^{(n)})$. If the speaker label for segment $n$ is $k$, then the quantity of $d_{nk}$ is 1, and 0 otherwise~\cite{Snyder2017DeepNN}. 

\begin{equation}
E = -\sum_{n=1}^{N} \sum_{k=1}^{K} d_{nk} \ln{P(spkr_k | x_{1:T}^{(n)})}.
\label{eq_entropy}
\end{equation}

\subsubsection{Embeddings}
The embeddings produced by the network described above capture information from the speakers over the whole audio-signal.
 Such embeddings are called \textit{x-vectors} and they can be extracted from any \textit{segment} layer; that is, either \textit{segment6} or \textit{segment7} layers (see Table~\ref{table_dnn}). Normally, embeddings from the \textit{segment6} layer give a better performance than those from \textit{segment7}~\cite{snyderxvecs}. In this study, these type of representations can capture meaningful information from each utterance. This embedding may help us to discriminate better the utterances due to the fact that the characteristics are acquired at the utterance level rather than at the frame-level. For this, we used the Kaldi Toolkit~\cite{kaldi}.

\subsection{Fisher Vectors}
The Fisher Vector approach is an image representation that pools local image descriptors (e.g. SIFT, describing occurrences of rotation- and
scale-invariant primitives~\cite{lowe2004distinctive}). In contrast with the Bag-of-Visual-Words (BoV,~\cite{peng2016bag}) technique, it assigns a local descriptor to elements in a visual dictionary, obtained via a Gaussian Mixture Model for FV. Nevertheless, instead of just storing visual word occurrences, these representations take into account the difference between dictionary elements and pooled local features, and they store their statistics. A nice advantage of the FV representation is that, regardless of the number of local features (i.e. SIFT), it extracts a {\it fixed-sized} feature representation from each image.

The FV technique has been shown to be quite promising in image representation~\cite{jaakkola1998exploiting}. Despite the fact that just a handful of studies use FV in speech processing (e.g. for categorizing audio-signals as speech, music and others~\cite{moreno2010usingthefisher}, for speaker verification~\cite{tian2014speakerverification,zajic2016fishervectors}, and for determining the food type from eating sounds~\cite{kaya2015fishervectors}), we think that FV can be harnessed to improve classification performance in audio processing. 

\subsubsection{Fisher Kernel}
The Fisher Kernel (FK) seeks to measure the similarity of two objects from a parametric generative model of the data ($X$) which is defined as the gradient of the log-likelihood of $X$:
\begin{equation}
G_{\lambda}^{X}= \bigtriangledown_{\lambda}\log\upsilon_{\lambda}(X),
\label{eq_glambda}
\end{equation}
where $X=\{x_t, t=1,\ldots, T\}$ is a sample of $T$ observations $x_t \in {\cal X}$, $\upsilon$ represents a probability density function that models the generative process of the elements in ${\cal X}$ and $\lambda=[ \,\lambda_1,\ldots,\lambda_M]\,' \in R^M$ stands for the parameter vector $\upsilon_\lambda$~\cite{sanchez2013imageclassification}. Thus, such a gradient describes the way the parameter $\upsilon_\lambda$ should be changed in order to best fit the data $X$. A novel way to measure the similarity between two points $X$ and $Y$ by means of the FK can be expressed as follows~\cite{jaakkola1998exploiting}:
\begin{equation}
K_{FK}(X,Y) = G_{\lambda}^{X\prime} F_{\lambda}^{-1} G_{\lambda}^{Y}.
\label{eq_kfk}
\end{equation}
Since $F_\lambda$ is positive semi-definite, $F_\lambda=F_\lambda^{-1}$. Eq.~(\ref{eq_dotkfk}) shows how the Cholesky decomposition $F^{-1}_\lambda=L_\lambda' L_\lambda$ can be utilized to rewrite the Eq.~(\ref{eq_kfk}) in terms of the dot product:
\begin{equation}
K_{FK}(X,Y) = \mathscr{G}_{\lambda}^{X\prime} \mathscr{G}_{\lambda}^{Y},
\label{eq_dotkfk}
\end{equation}
where
\begin{equation}
\mathscr{G}_{\lambda}^{X} = L_\lambda G_{\lambda}^{X} = L_\lambda \bigtriangledown_{\lambda}\log\upsilon_{\lambda}(X).
\label{eq_normG}
\end{equation}
Such a normalized gradient vector is the so-called \textit{Fisher Vector} of $X$~\cite{sanchez2013imageclassification}. Both the FV $\mathscr{G}_{\lambda}^{X}$ and the gradient vector $G_{\lambda}^{X}$ have the same dimension.

\subsubsection{Fisher Vectors}

Let $X=\{X_t, t=1\ldots T\}$ be the set of $D$-dimensional local SIFT descriptors extracted from an image and let the assumption of independent samples hold, then Eq.~(\ref{eq_normG}) becomes:
\begin{equation}
\mathscr{G}_{\lambda}^{X} = \sum_{t=1}^{T} L_\lambda \bigtriangledown_{\lambda}\log\upsilon_{\lambda}(X_t).
\label{eq_sumLambda}
\end{equation}
The assumption of independence permits the FV to become a sum of normalized gradients statistics $L_\lambda \bigtriangledown_\lambda \log\upsilon_\lambda(x_t)$ calculated for each SIFT descriptor. That is:
\begin{equation}
X_t\to \varphi_{FK}(X_t) = L_\lambda \bigtriangledown_{\lambda}\log\upsilon_{\lambda}(X_t),
\label{eq_higEmb}
\end{equation}
which describes an operation that can be thought of as a higher dimensional space embedding of the local descriptors $X_t$.

The FV extracts low-level local patch descriptors from the audio-signal spectrogram. Then, a GMM with diagonal covariances models the distribution of the extracted features. The log-likelihood gradients of the features modeled by the parameters of such GMM are encoded through the FV \cite{sanchez2013imageclassification}. This type of encoding stores the mean and covariance deviation vectors of the components $k$ that form the GMM together with the elements of the local feature descriptors. The utterance is represented by the concatenation of all the mean and the covariance vectors that gives a final vector of length $(2D+1)N$, for $N$ quantization cells and $D$ dimensional descriptors~\cite{sanchez2013imageclassification, perronnin-fisherkernels}. Here, we use FV features to encode the MFCC features extracted from the audio-signals of the \textit{Mask} dataset. 

\subsection{Support Vector Machines (SVM)}
A linear-SVM classifier was utilized to discriminate the audio-signals. This algorithm was found to be robust even with a large number of dimensions and it was shown to be efficient when used with FV~\cite{sanchez2013imageclassification,smith2013comparison} due to it being a discriminative classifier that provides a flexible decision boundary. We used the libSVM implementation~\cite{libsvm} with a linear kernel. As stated in the paper on this year's challenge~\cite{compare2020}, since 2009 (and also for this year), Unweighted Average Recall (UAR) has been the chosen metric for evaluating the performance of the classifiers.

\section{Experimental setup}
As for the Fisher vectors, the number of GMM components $K$ utilized to compute the FVs ranged from $2, 4, 8,$ to $512$. The construction of the FV encoding was performed using a Python-wrapped version of the VLFeat library~\cite{vedaldi2010vlfeat}. Both MFCC and PLP representations were used separately to train the GMM model and extract the FV features. The GMM model was fit utilizing the training set. Fisher vectors were \textit{optimized} employing Power Normalization (PN) and L2-Normalization before training the data; in~\cite{sanchez2013imageclassification} the authors show obtained good FV performance using this feature pre-processing technique.

The x-vector network was fitted using the training data and its augmented version following the methodology employed in~\cite{snyderxvecs}; likewise, we used the same network topology proposed there. Basically, from the original training data, two augmented versions were added, i.e. noise and reverberation. From additive noises and reverberation, two of the following types of augmentation were chosen arbitrarily: babble, music, noise, and reverberation. The first three types correspond to simply adding or fitting a kind of noise to the original utterances, while the fourth one involves a convolution of room impulse responses with the audio, i.e. reverberation~(see \cite{snyderxvecs} for more details about the augmentation strategies used). From the artificially generated data, we chose a subset of 40000 utterances to train the DNN, which is roughly four times the number of original training samples. From the \textit{segment6} layer of the DNN, we extracted 512-dimensional neural network embeddings (x-vectors) for the train, development, and test sets, respectively. As Snyder et al.~\cite{snyderxvecs}, we also found that embeddings from \textit{segment6} gave a better performance than those from \textit{segment7} in our experiments.

Following the techniques suggested in~\cite{jaakkola1998exploiting}, the parameter $C$ of the SVM was set in the range: $10^{-5}$, $\ldots$, $10^1$. Since the training and development sets are meant to be combined and used to train the final SVM model, we fused the above-mentioned sets and employed a Stratified \textit{k}-fold Cross-Validation. We set $k=10$ to find the best $C$. The training set has 5353 utterances labeled as \textit{no-mask} and 5542 labeled as \textit{mask}; the development set has 6666 and 7981, as \textit{no-mask} and as \textit{mask}, respectively. Namely, there is a slight class imbalance when combining both sets. As a result, there were 1504 more utterances labeled as \textit{mask} in the combin\-ed set. Hence, we set the \textit{class-weight} parameter of the SVM to \textit{balanced}. In this way, the classifier adjusted the weights of the classes automatically. Before classification, all the features were standardized by removing their means and scaling to unit variance.

In addition, we carried out a \textit{late fusion} of our best configurations. Moreover, we also fused our best configurations with those posteriors from \textit{`fusion of best'} of the sub-challenge~\cite{compare2020}. 

\begin{table}[t!]
\centering
\caption{Experiment results. Scores are presented for x-vectors and FVs; both using MFCCs and PLPs. FoB stands for `fusion of best' (fusion of the ComParE best configurations)~\cite{compare2020}.
The GMM size corresponds to the $K$ value used for FV; for x-vectors this is not applicable. The dashes (-) in the UAR column indicate that the scores for those configurations are not available due to the limited number of trials for submissions defined by the organizers of the Challenge.}\label{tab_scores1}
\begin{tabular}{@{}lcccc@{}}
\toprule
\multicolumn{1}{c}{\multirow{2}{*}{\textbf{Feature}}} &
  \multirow{2}{*}{\textbf{\begin{tabular}[c]{@{}c@{}}GMM \\ size\end{tabular}}} &
  \multicolumn{3}{c}{\textbf{UAR (\%)}} \\ \cmidrule(l){3-5} 
\multicolumn{1}{c}{}     &       & \textbf{Dev} & \textbf{CV} & \textbf{Test} \\ \midrule
$ComParE \ Baseline_{FoB}$      & -     & -            & -           & 71.8          \\ \midrule
$x$-$vecs_{MFCC}$              & -     & 56.86        & 65.21       & -             \\
$x$-$vecs_{MFCC-hires}$        & -     & 59.87        & 72.14       & -             \\
$x$-$vecs_{PLP}$              & -     & 58.46        & 64.8         & -          \\
$FV_{MFCC}$                  & 512   & 57.43        & 78.18        & -             \\
$FV_{PLP}$                   & 256   & 59.18        & 71.09        & -             \\ \midrule
$FV_{512}$ + $FoB$  & -     & -            & -           &70.3               \\
$FV_{512}$ + $x$-$vecs_{hires}$                & - & -            & -           & 70.8          \\
$FV_{512}$ + $x$-$vecs_{hires}$ + $FoB$ & - & -            & -           &74.9  \\      
 \bottomrule
\end{tabular}
\end{table}

\section{Results and discussion}
As Table~\ref{tab_scores1} shows, the FV representations produced better performances in the evaluation (i.e. Dev and CV) phase than the x-vectors embeddings did. However, this is mainly true for the CV scores, where FV achieved UAR scores above 70\%. Overall, the configuration \textit{FV (MFCC)} attained the best CV score (78.18\%). On the other hand, the best configuration for the x-vectors embeddings was that of \textit{high resolution} MFCCs (i.e. MFCC-hires), which gave a 72.14\%. In contrast, when we evaluated the features using just the development set, x-vectors presented better scores; nevertheless, the difference compared to those of FV was not significant.
Although in this study we did not rely on the development scores to find the best $C$ value for the SVM, we still report the scores obtained when evaluating on this dataset (see Table~\ref{tab_scores1}). It should be added that we chose the best $C$ based on the Stratified 10-fold CV experiments.

Furthermore, the FV encodings yielded a significant performance improvement when applying PN and L2-normalization before fitting them (see also~\cite{sanchez2013imageclassification}). However, here, just the best configurations are reported (the improved FVs). PN reduced the effect of the features that become more sparse as the value of $K$ increased. Also, L2-normaliza\-tion helped to alleviate the problem of having different utterances with (relatively) distinct amounts of backgrou\-nd information projected into the extracted features. This mainly enhanc\-ed the prediction performances. Also, it was found that the higher the number of $K$, the higher the UAR score. This means that these two are directly proportional to each other.
In our study, both MFCC and PLP achieved their best configurations when using a large value for $K$ (512 and 256, respectively). Likewise, MFCC-hires gave a better frame-level feature quality (for the x-vectors) than the standard MFCC configuration. This can be attributed to the DNN training phase, where the neural network exploits in a better way the larger and less correlated frame-level representations.

Table~\ref{tab_scores1} lists the final scores. The fusion of the posteriors of \textit{$FV_{512}$} with those of the \textit{fusion of best} (from the challenge) attained a UAR score of 70.3\% on the test set.
Likewise, the fusion of \textit{$FV_{512}$} with x-vectors \textit{($x$-$vecs_{hires}$)} yielded a score of 70.8\%. Finally, the fusion of \textit{$FV_{512}$} with \textit{$x$-$vecs_{hires}$} along with \textit{FoB} provided a UAR score of 74.9\% on the test set.

\section{Conclusions}
Here, we studied the performance of x-vector and Fisher vector representations as a contribution to the \textit{Mask} Sub-Challenge of the INTERSPEECH 2020 ComParE. These representations were extracted from two different types of frame-level features: MFCC and PLP. As for the FV encodings, we found that MFCCs presented a superior type of frame-level traits of the recordings than the PLP did. Regarding the x-vectors, the configuration of \textit{MFCC-hires} was found to be better than those of the standard MFCC and PLP.
Also, we found that PN and L2-Normalizat\-ion enhanced the quality of the FVs. Although the FV gave better quality features than x-vectors for this particular dataset, x-vectors also captured meaningful phonatory with articulatory information, as their scores are competitive.
Moreover, we found that the fusion of our best configurations increased the performance of the final predictions. To conclude, our workflow outperformed the official baseline scores of the \textit{Mask} Sub-Challenge~\cite{compare2020}; besides, our feature extraction approach appears to be simpler than those from~\cite{compare2020}.

\bibliographystyle{IEEEtran}

\bibliography{mybib}

\end{document}